\documentclass[prl,aps,twocolumn,superscriptaddress,longbibliography]{revtex4-2}
\DeclareUnicodeCharacter{2212}{\textminus}
\usepackage{epstopdf}
\usepackage{lineno}
\usepackage{epsfig}
\usepackage{hyperref}
\usepackage{amsmath}

\usepackage{url}
\usepackage[caption=false,font={normalsize}]{subfig}
\usepackage{graphicx}
\usepackage{verbatim}
\usepackage{multirow}
\usepackage{amsmath, bm}
\newcommand{\beq}{\begin{eqnarray}}
\newcommand{\eeq}{\end{eqnarray}}
\usepackage{cleveref}
\usepackage{mathrsfs}
\usepackage{float}
\usepackage[usenames, dvipsnames]{color}
\usepackage{mathtools}
\usepackage{slashed}
\usepackage{physics}	
\usepackage{graphicx}   
\usepackage{epstopdf}
\usepackage{hyperref}   
\hypersetup{
pdfnewwindow=true, colorlinks=true,
linkcolor=blue, anchorcolor=blue,
citecolor=blue, filecolor=blue,
menucolor=blue, urlcolor=blue} 

\usepackage[export]{adjustbox}

\usepackage{bbold}
\usepackage{wasysym}
\usepackage{amssymb}
\definecolor{darkgray}{RGB}{10,30,30}
\definecolor{blue2}{rgb}{0.2, 0.2, 0.6}
\definecolor{blue3}{rgb}{0.16, 0.32, 0.75}
\definecolor{darkred}{rgb}{0.8,0,0}
\definecolor{royalblue}{rgb}{0.0, 0.14, 0.4}
\definecolor{magenta}{cmyk}{0,.9,0,0.2}
\definecolor{amethyst}{rgb}{0.6, 0.4, 0.8}
\definecolor{cadmiumgreen}{rgb}{0.0, 0.42, 0.24}
\definecolor{deepcarmine}{rgb}{0.66, 0.13, 0.24}
\definecolor{forestgreen}{rgb}{0.13, 0.55, 0.13}
\newcommand{\ncmd}{\newcommand}
\ncmd{\sect}[1]{\emph{\textbf{{#1}}}~---~}

\ncmd{\para}[1]{\paragraph*{{\color{black}{\bf #1:}}} }

\ncmd{\note}[1]{{\color{gray}{[\ding{168} #1}]}}

\ncmd{\YWnote}[1]{{\color{purple}{[\ding{168} {\bf #1}}]}}

\ncmd{\sur}[1]{{\color{forestgreen}{ #1}}}
\ncmd{\qs}[1]{{\color{red}{ #1}}}
\ncmd{\qsnote}[1]{{\color{cyan}{ #1}}}

\newcommand{\beginsupplement}{
        \setcounter{table}{0}
        \renewcommand{\thetable}{S\arabic{table}}
        \setcounter{figure}{0}
        \renewcommand{\thefigure}{S\arabic{figure}}
        \setcounter{equation}{0}
        \renewcommand{\theequation}{S\arabic{equation}}
        \setcounter{section}{0}
        \renewcommand{\thesection}{\Alph{section}}
        \setcounter{subsection}{0}
        \renewcommand{\thesubsection}{\arabic{subsection}}
}

\ncmd{\yw}[1]{{\color{blue}{#1}}}

\begin{document}
\preprint{}

\title{Kondo destruction quantum critical point: fixed point annihilation and thermodynamic stability}
\author{Yiming Wang}
\affiliation{Department of Physics \& Astronomy,  Extreme Quantum Materials Alliance, Smalley-Curl Institute, Rice University, Houston, Texas 77005, USA}
\author{Lei Chen}
\affiliation{Department of Physics \& Astronomy,  Extreme Quantum Materials Alliance, Smalley-Curl Institute, Rice University, Houston, Texas 77005, USA}
\author{Haoyu Hu}
\affiliation{Department of Physics \& Astronomy,  Extreme Quantum Materials Alliance, Smalley-Curl Institute, Rice University, Houston, Texas 77005, USA}
\author{Ang Cai}
\affiliation{Department of Physics \& Astronomy,  Extreme Quantum Materials Alliance, Smalley-Curl Institute, Rice University, Houston, Texas 77005, USA}
\author{Jianhui Dai}
\affiliation{School of Physics, Hangzhou Normal University, Hangzhou 310036, China}
\affiliation{Department of Physics \& Astronomy,  Extreme Quantum Materials Alliance, Smalley-Curl Institute, Rice University, Houston, Texas 77005, USA}
\author{C. J. Bolech}
\affiliation{Department of Physics, University of Cincinnati, 345
Clifton Court, Cincinnati, Ohio 45221-0011, USA} 
\affiliation{Department of Physics \& Astronomy,  Extreme Quantum Materials Alliance, Smalley-Curl Institute, Rice University, Houston, Texas 77005, USA}
\author{Qimiao Si}
\affiliation{Department of Physics \& Astronomy,  Extreme Quantum Materials Alliance, Smalley-Curl Institute, Rice University, Houston, Texas 77005, USA}
\date{\today}
\begin{abstract}
A wide range of strongly correlated electron systems exhibit strange metallicity, and they are 
 increasingly recognized as in proximity to correlation-driven localization-delocalization transitions. A prototype setting arises in heavy fermion metals, where the proximity to the electron localization is manifested as Kondo destruction. Here we show that the Kondo destruction quantum critical point is linked to the phenomenon of fixed point annihilation. This connection reveals the absence of residual entropy density at the quantum critical point and, thus, its thermodynamic stability. Broader implications of our results are discussed.
\end{abstract}

\maketitle

\sect{Introduction} 
Strange metallicity develops in a variety and expanding list of strongly correlated electron systems~\cite{Keimer-Moore_2017,Paschen-Si_2020,Phillips_stranger_2022,hu_quantum_2024}. In addition to canonical systems of correlated electrons, such as the cuprates~\cite{lee_doping_2006} and heavy fermion metals~\cite{Paschen-Si_2020,kirchner_colloquium_2020}, it has been observed in moiré and other flat band systems~\cite{checkelsky_flat_2024}. An emerging profile of strange metals includes, along with a linear-in-temperature electricial resistivity, dynamical Planckian ($\hbar \omega/k_{\rm B}T$) scaling, a jump of Fermi surface, and loss of quasiparticles. 
These properties suggest that strange metallicity develops through proximity of the system to a correlation-driven localization-delocalization transition. The case is particularly striking in heavy fermion strange metals, where such a transition corresponds to the localization of the Kondo-induced quasiparticles, or Kondo destruction
\cite{si_locally_2001,coleman_how_2001,senthil_weak_2004}. Extensive experimental measurements in heavy fermion strange metals have provided strong evidence for the Kondo destruction description \cite{Paschen-Si_2020,kirchner_colloquium_2020,Gegenwart_Si_Steglich}. These include
the inelastic neutron scattering spectra~\cite{Aronson_1995,schroder_onset_2000,Mazza2024Quantum}, optical conductivity~\cite{prochaska_singular_2020},  magnetotransport~\cite{paschen_hall-effect_2004,gegenwart_multiple_2007,friedemann_fermi-surface_2010,Custers_2012,Martelli2019} and quantum oscillation measurements~\cite{shishido_drastic_2005}, as well as measurements of the Wiedemann-Franz ratio~\cite{pfau_thermal_2012} and shot noise Fano factor~\cite{chen_shot_2023}.

Kondo destruction quantum criticality appears in Kondo lattice models, which contain $f$-electrons in the form of localized spin-$1/2$ moments and itinerant
electrons~\cite{Hewson}. Antiferromagnetic Kondo interaction between the local moments and conduction electrons favors the formation of a Kondo-singlet ground state, from which the local moments are converted into heavy quasiparticle excitations; 
whereas the RKKY interaction among the local moments promotes magnetic order or related ground states of quantum magnetism. Within an extended dynamical mean field theory (EDMFT) approach
\cite{hu_quantum_2024,si_locally_2001,Si1-2,Smith-2,Chitra2000}, this competition appears through a Bose-Fermi Kondo model (BFKM), which describes the coupling of a local moment to both a fermionic bath and a bosonic one. While the fermionic bath has a nonzero density of states at the Fermi energy, the bosonic bath has the following power-law spectrum:
\begin{eqnarray}
J(\omega) \propto |\omega|^{1-\epsilon} ~\mathrm{sgn}(\omega)
~\, .
\label{dissipative_spectrum}
\end{eqnarray}
Here the exponent $\epsilon$ characterizes the spectrum of the 
fluctuating magnetic field that a local moment spin experiences.
At the Kondo desttruction quantum critical point,
the EDMFT self-consistency between the correlation functions of the 
local moment spin and the spectrum of the effective fluctuating magnetic field  
dictates that the exponent $\epsilon$ takes a particular value \cite{si_locally_2001,Grempel,Glossop07,Zhu07}:
\begin{eqnarray}
\epsilon \,=\, 1^{-}
~\, .
\label{epsilon-KD}
\end{eqnarray}

\begin{figure}[t!]
    \centering
    \includegraphics[width=\linewidth]{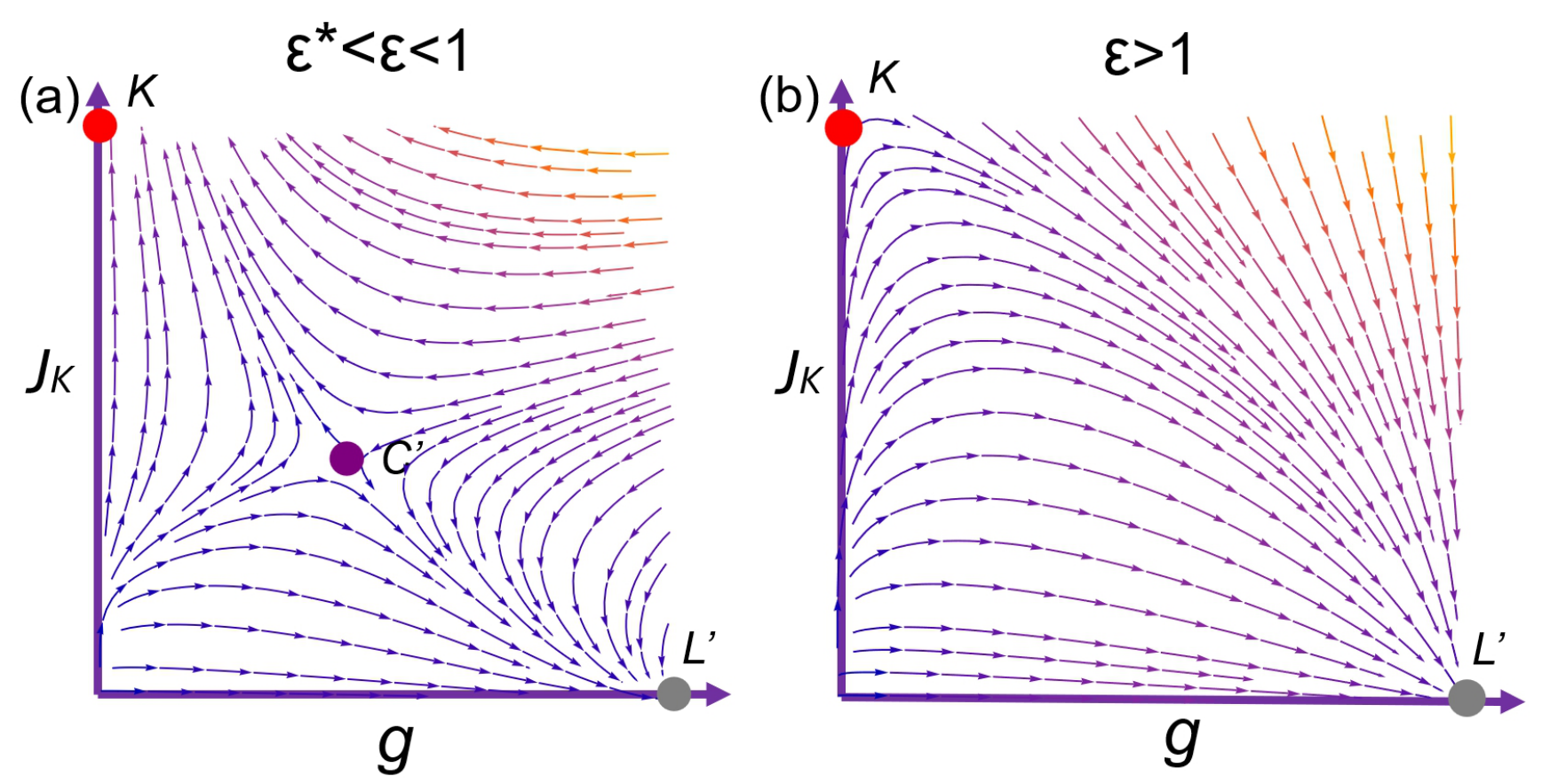}
    \caption{Schematic RG flow of the SU(2) symmetric Bose-Fermi Kondo model for (a) $\epsilon^*<\epsilon < 1$, 
    with $\epsilon^*=0.53$, 
    and (b) $\epsilon > 1$. The red dot 
$K$ denotes the Kondo fixed point, the purple dot $C'$
 represents the Kondo destruction quantum critical point (QCP), and the gray dot $L'$
 corresponds to the local moment fixed point.}
    \label{fig:RG-illustration}
\end{figure}

Importantly, in this approach, the Kondo destruction strange metal state is associated with a renormalization-group (RG) fixed point. This association provides the opportunity to address some of the key theoretical questions on the nature of the state. Here we harness this linkage to demonstrate the thermodynamic stability of the Kondo destruction quantum critical state. We show that the quantum critical state does not possess any zero temperature residual entropy and, thus, is thermodynamically stable. As illustrated in Fig.\,\ref{fig:RG-illustration}(a,b), 
we have done so by connecting the underlying quantum critical state with the phenomenon of fixed point annihilation. The latter refers to the topology of the RG flow generally permitting the collision of two fixed points differing by one in the number of relevant directions. As the flow parameters are varied, the two fixed points can move toward one another in the coupling space, eventually merging and disappearing. This mechanism of 
fixed-point annihilation, discussed early on in the context of QCD-like theories \cite{kaplan}, have gradually emerged 
in diverse contexts, including the non-linear sigma model \cite{adam} and models for boundary critical phenomena \cite{hu2022,cuomo_spin_2022,adam2}.

\sect{Bose-Fermi Kondo model at $0<\epsilon \le 2$}
Our strategy is to provide an understanding of the BFKM as $\epsilon$ is lowered from $2$. This approach reveals a perspective that connects the Kondo destruction QCP at $\epsilon=1^{-}$ to the notion of fixed point annihilation.

The 
Hamiltonian for 
the BFKM is
\begin{eqnarray}
H=H_{F}+H_{B}+H_{K}+H_{g}+H_h \, .
\label{hamiltonian_bfkm}
\end{eqnarray}
Here, $H_F=\sum_{k,\sigma}\varepsilon_{k}c^{\dagger}_ {k\sigma}
c_{k\sigma}$ and 
$H_B=\sum_{p,a}w_{p}\Phi^{\dagger}_{pa}\Phi_{pa}$
respectively describes a conduction-electron band 
and a bosonic bath, with $\varepsilon_k$ and $w_{p}$ denoting 
the corresponding dispersions. $\sigma=\uparrow,\downarrow$ marks the spin quantum number,
and $a=1,2,3$ the components of the vector boson field. 
$H_K=(J_{K\perp}/2) [s^-_c(0)S^+
+s^+_c(0)S^-]+J_{K||}s^z_c(0)S^z$ 
specifies the Kondo coupling
between
a spin-$1/2$ local moment, $\vec{S}$,
and the conduction electron spin
$\vec{s}_c=\sum\frac{1}{2}c^{\dagger}_{k\sigma}\vec{\tau}_{\sigma,\sigma'}
c_{k\sigma'}$ (with $\vec{\tau}$ being the Pauli matrices) at the spin site $0$
and $H_g=\sum_{a=1}^3g_a\Phi^aS^a$ 
the counterpart coupling between the spin and the bosonic field,
with $\Phi^a \equiv 
\sum_{p}(\Phi_{pa}+\Phi^{\dagger}_{-pa})$.
The dissipative bosonic spectrum,
$J(\omega)
\equiv \sum_{p}[\delta(\omega-w_{p})-\delta(\omega+w_{p})]$,
is
taken to have the form of Eq.\,(\ref{dissipative_spectrum}). 
We will only be concerned with the sub-ohmic case,  corresponding to $\epsilon>0$.
In order to probe the
spin responses, we
introduce a local magnetic field, $h$, 
via $H_h=hS^z$.
The BFKM was first introduced in Ref.\,\onlinecite{Smith-2}, where the focus 
was on the spin-anisotropic case and where  
an $\epsilon$-expansion
RG analysis was carried out.
Subsequent studies for the spin-isotropic case were 
performed in Refs.\,\onlinecite{Smith,Sengupta}.
The Kondo coupling to the fermionic bath {\it per se} is standard \cite{Hewson}, 
whereas that to the bosonic bath {\it per se} appears in dissipative quantum mechanics 
and other contexts \cite{Leggett,Bray-Moore,Sachdev}. 
RG studies were carried out in the $\epsilon$ expansion to higher orders 
\cite{Zhu,Zarand}. Numerical studies 
used numerical renormalization group (NRG)~\cite{Glossop,Glossop_prb07,Pixley2013} and continuous time quantum Monte Carlo (CT-QMC) means~\cite{Pixley2013,cai2019bose}. The model also appears in the more recent studies of random $t-J$ Hamiltonians~\cite{Cha2020,Joshi2020,Dumitrescu2022}.
Finally, BFKM is furthermore of interest in the context of
dissipative effects
in mesoscopic structures~\cite{Hur,Kirchner}.

The partition function of the BFKM
can be written in the path-integral
form.
The $s=1/2$ local moment spin, $\vec{S}$, is
represented by an $SU(2)$ coherent state $S\vec{\Omega}(\tau)$
with
the constraint $\vec{\Omega}^2(\tau)=1$. A Berry phase term
$\mathcal{S}_{WZ}[\vec{\Omega}]$ characterizes the quantum dynamics
of the spin. Meanwhile, the bosonic bath can be traced out,
leading to a long-ranged interaction 
(along the $\tau$-dimension) for the  spin:
\begin{eqnarray}
Tr
\exp\{-\beta
(H_B+H_g)\}=Z_b
\int {\cal D}\vec{\Omega}
\exp\{-i
\mathcal{S}_{WZ}[\vec{\Omega}]\}~~~~\\
\times \exp\{\sum_{a=1}^3\int_0^{\beta}\int_0^{\beta}
d\tau_1d\tau_2 {\cal L}(\tau_1,\tau_2)
\}\, ,\nonumber\label{bkm}
\end{eqnarray}
where 
\begin{eqnarray}
{\cal L}(\tau_1,\tau_2) = 
\left [ {g^2_aS\Omega^a(\tau_1)S\Omega^a(\tau_2)/4} \right ]
\chi_{0}^{-1}(\tau_{2}-\tau_{1}) 
\, .\nonumber
\end{eqnarray}
Here, $Z_b$ is
the partition function of the bosonic bath, 
and $\chi_{0}^{-1}(\tau_{2}-\tau_{1})=1/{|\tau_2-\tau_1|^{2-\epsilon}}$.
(Note
that the purely classical model defined along a chain,
$0<\tau<\beta$, 
would be ill-defined for
$1<\epsilon \le 2$, since the long-ranged interaction makes 
the ground state energy per unit length diverge 
in the $\beta \rightarrow
\infty$ limit, but the quantum problem we consider is 
well-defined \cite{note}).

We will pay a particular attention to $\epsilon=2$.
Here, the long-range interaction in the
last term becomes $\sum_{a=1}^3[\int_0^{\beta} d\tau
g_aS\Omega^a(\tau)/2]^2$, which can be further decomposed by
introducing a Hubbard-Stratanovich vector
$\vec{\lambda}=(\lambda_1,\lambda_2,\lambda_3)$:
\begin{eqnarray}
\exp\{\sum_{a=1}^3[\int_0^{\beta} d\tau
Sg_a\Omega^a(\tau)/2]^2\}=~~~~~~~~~~~~~~~~~~~\nonumber\\
\pi^{-3/2}\int d\vec{\lambda}\exp\{-\vec{\lambda}^2-\sum_{a=1}^3
g_a\lambda_a\int_0^{\beta}d\tau S\Omega^a(\tau)\} \, .
\end{eqnarray}
The total partition function of the BFKM, when rewritten in the
operator formalism, becomes
\begin{eqnarray}
Z=\mathrm{Tr} e^{-\beta H}&=&\pi^{-3/2}Z_b\int d\vec{\lambda}
e^{-\vec{\lambda}^2} \mathrm{Tr}
e^{-\beta \tilde{H}(\vec{\lambda})}\, ,\nonumber\\
\tilde{H}(\vec{\lambda})&=&H_F+ H_K+H_h+H_{\vec{\lambda}} \, ,
\label{Z_gaussian_averaging}
\end{eqnarray}
where $H_{\vec{\lambda}}=\sum_{a=1}^3g_a\lambda_aS^a$. Hence, for $\epsilon=2$, we
are led to solve the BFKM, Eq.~(\ref{hamiltonian_bfkm}),
in terms of a pure Fermi-Kondo model in the presence of a
Gaussian-distributed magnetic field $\vec{\lambda}$,
along
with the external field $h$.

\sect{Solution for $J_K=0$ at $\epsilon \le 2$}
We first consider the case with $g_a=g$ (for $a=1,2,3$) and 
without the conduction electrons, {\it i.e.}, the $SU(2)$ Bose-Kondo model.
At $\epsilon=2$, the 
partition function $Z_{loc}= Z_b^{-1}Z$ can be solved analytically 
[See the Supplementary Material (SM), Sec.\,~\ref{Sec:A}], 
leading to the following result for the static local spin susceptibility:
\begin{eqnarray}
\chi_{loc}
=\frac{\beta}{12}\frac{3+{g^2\beta^2}/{8}}{1+{g^2\beta^2}/{8}} 
~\,\stackrel{\beta \rightarrow \infty}{\longrightarrow}\,
~\frac{\beta}{12} \, .
\label{chi_loc_su2_static}
\end{eqnarray}
The asymptotic behavior in the low-temperature limit has a Curie form,
with a reduced Curie constant
($1/12$ instead of the free-spin value, $1/4$).

The dynamical local spin-spin correlation function,
$\chi_{loc}(\tau) \equiv \langle S^z(\tau)S^z(0)\rangle_{loc}$,
can be obtained from a Gaussian averaging (SM, Sec.\, \ref{Sec:A}).
We find
\begin{eqnarray}
\chi_{loc}(\tau)
&=&\frac{1}{12}\left [1+
\frac{2+\frac{g^2(\beta-2\tau)^2}{4}}{1+\frac{g^2\beta^2}{8}}
e^{-\frac{g^2}{4}\tau(\beta-\tau)} \right ]
\nonumber \\
&& \mathop{\longrightarrow}_{\beta \rightarrow \infty}^{\tau \rightarrow
\beta/2} ~\frac{1}{12} \, .
\label{chi_loc_su2}
\end{eqnarray}
In the asymptotic low-temperature and long-time limit
($\tau\rightarrow \beta/2$, $\beta \rightarrow \infty$),
it also has a Curie form.

\begin{figure}[t!]
\centering
  \mbox{\includegraphics[width=.90\columnwidth]{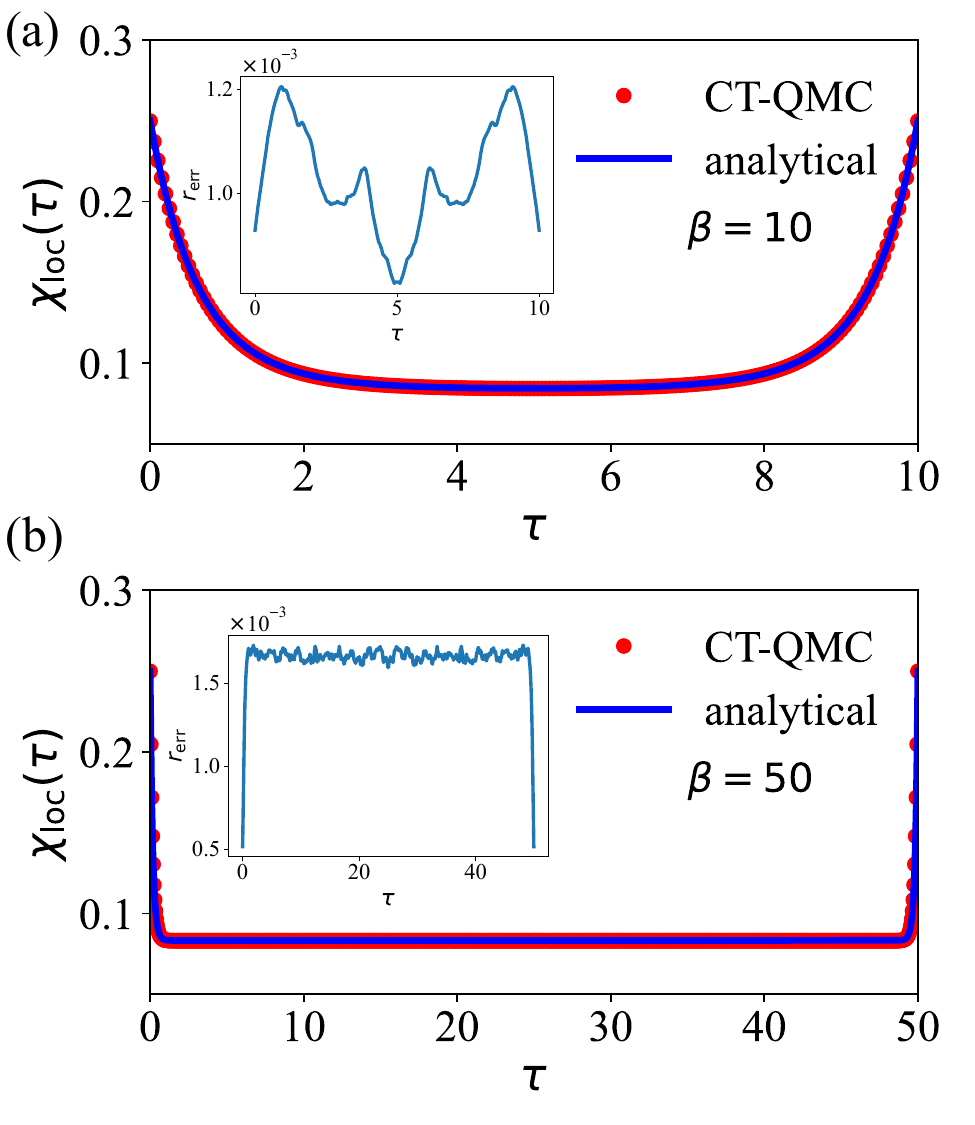}}    
\caption{
The dynamical local spin susceptibility of the $SU(2)$ Bose-only Kondo model at $\epsilon=2$ obtained from the anlytical result, Eq.\,(\ref{chi_loc_su2}),
and from CT-QMC calculations with $g^{2}=0.5$ at (a) $\beta=10$ and (b) $\beta=50$. 
The insets show the relative errors.
\label{fig:dysus}
}
\end{figure}

The model can also be studied numerically using 
the CT-QMC method \cite{cai2019bose,otsuki2013spin,Werner2007}(See the SM, Sec.\,\ref{Sec:B}). 
In Fig.\,\ref{fig:dysus}, we compare the analytical result of Eq.\,(\ref{chi_loc_su2}) 
with the numerical result for $g^{2}=0.5$ at two different inverse temperatures $\beta=10$ and $\beta=50$. 
We see that they agree well 
with each other (with a relative error that is on the order of $10^{-3}$, as seen in the insets).

\begin{figure}[t!]
\centering
  \mbox{\includegraphics[width=0.85\columnwidth]{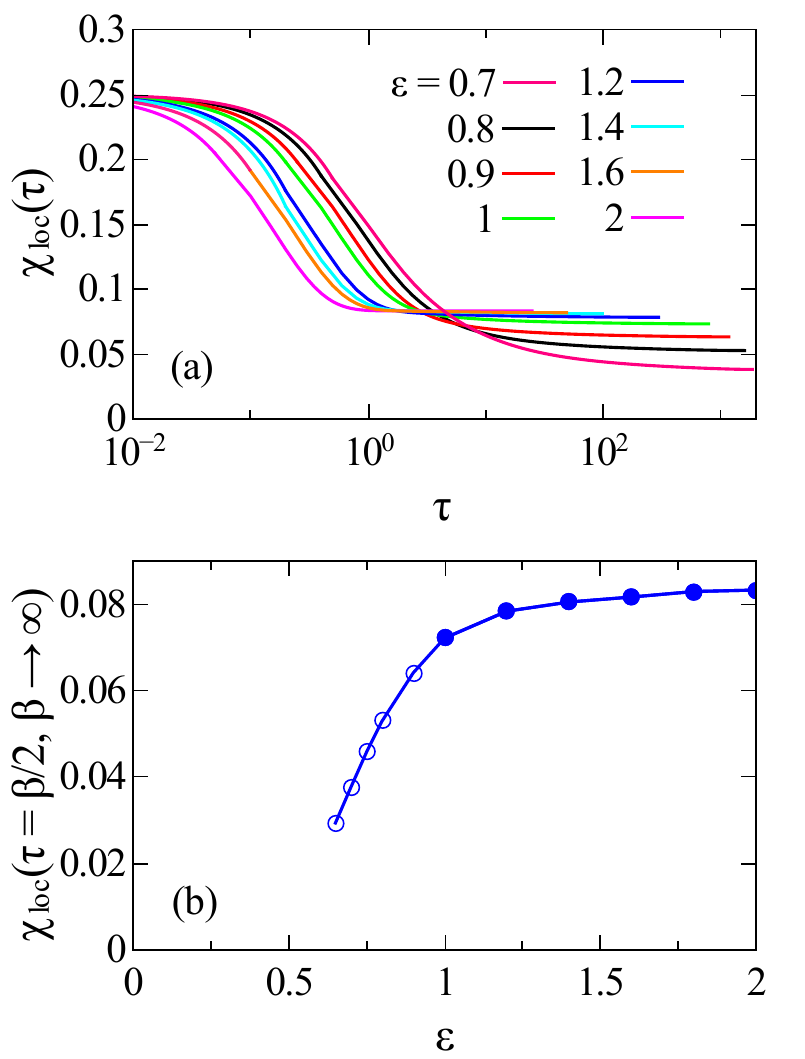}}    
\caption{  
(a) The dynamical local spin susceptibility, $\chi(\tau)$, from $\tau=0$ to $\tau=\beta/2$ calculated at $g^{2}=0.5$ and for different values of $\epsilon$.
(b) Effective Curie constant $\chi(\tau=\beta/2, \beta \rightarrow \infty)$ as a function of $\epsilon$, obtained at $g^{2}=0.5$. Filled points correspond to the $\epsilon \geq 1$ case calculated in this work. Empty points correspond to the cases of $\epsilon < 1$ \cite{cai2019bose}.
\label{fig:chi_vs_epsilon}
}
\end{figure}

The $\epsilon=2$ result anchors the understanding of the model at $\epsilon<2$. We continue to focus on the $J_K=0$ case. 
Fig.\,\ref{fig:chi_vs_epsilon}(a) shows $\chi_{loc}(\tau)$ from $\tau=0$ to $\tau=\beta/2$ obtained from CT-QMC 
at the largest possible $\beta$, for different value of $\epsilon$, from $\epsilon=2$ to $\epsilon=0.7$ (according to CT-QMC~\cite{cai2019bose}, the local moment fixed point L$^{\prime}$ that gives $\chi_{loc}(\tau)\sim 1/\tau^{0}$ always exists, but when $\epsilon<\epsilon^{*} \simeq 0.53$, it will require a sufficiently large $g$  to access L$^{\prime}$). At each $\epsilon$, $\chi_{loc}(\tau)$ starts from the maximized value $1/4$ at $\tau=0$ and decreases to a constant value at $\tau=\beta/2$. The calculations are performed at large enough $\beta$ for $\chi_{loc}(\tau=\beta/2)$ to converge to its asymptotic value at $\beta\rightarrow \infty$. We can define $\chi_{loc}(\tau=\beta/2,\beta\rightarrow \infty)$ as an effective Curie constant, and approximate its value using $\chi_{loc}(\tau=\beta/2)$ obtained at the largest $\beta$. 
It is seen that the effective Curie constant is reduced as $\epsilon$ becomes smaller.

This can be visualized in Fig. \ref{fig:chi_vs_epsilon}(b), where we summarize our result 
for the approximated $\chi_{loc}(\tau=\beta/2,\beta\rightarrow \infty)$ 
at different $\epsilon$. We see that $\chi_{loc}(\tau=\beta/2, \beta \rightarrow \infty)$ 
decreases from the exact value $1/12$ at $\epsilon=2$ to smaller values at $\epsilon<2$. 
In particular, $\chi_{loc}(\tau=\beta/2, \beta \rightarrow \infty)$ is smooth across $\epsilon=1$ and remains finite for $\epsilon<1$. This suggests that the local moment fixed point $L^{\prime}$ found at $\epsilon<1$~\cite{cai2019bose} is
connected 
to the same $g=\infty$ fixed point at $\epsilon=2$.

For comparison, we  
also consider the 
Ising case,
{\it i.e.},
$g_1=g_2=0,\, g_3=g$.
The
partition function can be determined (SM, Sec.\,\ref{Sec:A}),
and it follows that  
both the static and dynamical local spin susceptibilities
have simple Curie forms:
$\chi_{loc} ={\beta}/{4}$,
and $\chi_{loc}(\tau)={1}/{4}$.

Back to the $SU(2)$ 
case, 
the fact that the local spin response of 
for $\epsilon^*<\epsilon<2$
has a Curie form in the low temperature limit,
implies that,  
as in the Ising case,
it too is controlled by the 
$g^*=\infty$ fixed point.
In other words, 
the longitudinal fluctuations
dominate over the transverse ones.
This is to be contrasted with the large-$N$ limit in an SU(N)-symmetric formulation 
of the model
\cite{Kircan.04}; in the large-$N$ approach, the longitudinal interaction
represents one component among the $N^2-1$ parts and, thus, it does not appear 
in the leading order of a $1/N$ expansion. A related distinction is that,
for $1<\epsilon \le 2$, 
the large-$N$ result, $\chi_{loc}(\tau) \sim 1/\tau^{\epsilon}$,
violates the 
Griffiths' bound~\cite{Griffiths},
which requires  
$\chi_{loc}(\tau)$ {\it not} to decay faster than
the interaction, $1/\tau^{2-\epsilon}$.
By contrast, 
our exact result for the $SU(2)$ problem at $\epsilon=2$,
$\chi_{loc}(\tau) \sim 1/\tau^{0}$,
satisfies the Griffiths' bound.

\sect{The Bose-Fermi Kondo model at $\epsilon \le 2$} 
We now incorporate the Kondo coupling as well. 
Armed with the above insight, here,
we first focus on 
the case with the pure fermionic Kondo part being placed at 
its Toulouse point~\cite{Hewson,Giamarchi2003,Gogolin2004} and, moreover,
the bosonic coupling is purely Ising.  
As shown below, our result 
applies to the spin-isotropic BFKM.

We again start from the case of $\epsilon=2$. At the Toulouse point, the model Eq.~(\ref{Z_gaussian_averaging})
can be mapped to a non-interacting spinless
resonant level model~\cite{Hewson,note2}.

\begin{figure}[ht]
\includegraphics[width=.7\columnwidth]{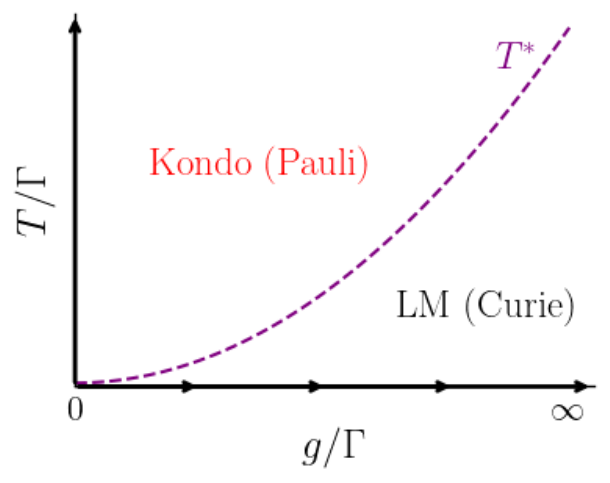}
\caption[]{The
phase diagram of the 
present BFKM.
$\chi_{loc}$ has the Curie or Pauli forms when $T\ll T^*$ or
$T^*\ll T\ll\Gamma$;
for $g \ll \Gamma$, 
$T^*\approx g^2/(2\pi\Gamma)$.
}
\label{fig:phase}
\end{figure}

At low temperatures, the zero-field static local susceptibility is obtained analytically (SM, Sec.\,\ref{Sec:C}).
The universal behavior is already captured for 
weak boson-spin couplings, $g/\Gamma \ll 1$ (where $\Gamma$ is the 
bare width of the resonant level). 
What emerges is a characteristic temperature scale, $T^*=\frac{
g^2}{2\pi\Gamma}$,
which separates two limiting temperature regimes with distinct asymptotic behavior of
the local spin susceptibility, $\chi_{loc}$.
At low temperatures, {\it i.e.}, $T\ll T^*$,
the local spin susceptibility shows a Curie behavior:
$\chi_{loc}\approx\frac{\beta}{4}$. 
In the higher temperature regime,
$T^* \ll T \ll \Gamma$,
the local spin susceptibility displays a Pauli behavior:
$\chi_{loc}\approx\frac{1}{\pi\Gamma}$. 
A
smooth but broad crossover occurs 
around $T \sim T^*$. This 
is illustrated in
Fig.~\ref{fig:phase}.

In other words, at $\epsilon=2$, for a fixed non-zero Kondo coupling (and, hence,
non-zero $\Gamma$), the local spin susceptibility in the
low-temperature limit turns to 
a Curie form  
for any non-zero $g$.
This implies that an infinitesimal $g$ causes a
destruction of the Kondo effect; 
the BFKM is
controlled by the fixed point associated with the Bose-only Kondo
model solved in the previous section.

This solution of the BFKM for $\epsilon=2$ again anchors the understanding of the model at $\epsilon<2$. 
The key is that, for $1 < \epsilon \le 2$,
the bosonic spectral density is singular as $\omega$ is reduced towards zero, {\it cf.} Eq.\,(\ref{dissipative_spectrum}).  
In this entire range of $\epsilon$, we can determine the 
fate of the Kondo state
in the presence of
a nonzero but arbitrarily small bosonic coupling $g$.

For this purpose, we consider 
the SU(2) Bose–Fermi Anderson model,
$g_a=g$, for $a=1,2,3$ [{\it cf.} Eq.\,(\ref{hamiltonian_bfkm})].
From
Firsov–Lang and Schrieffer–Wolff transformations \cite{Pixley_2011}  (SM Sec.\ref{Sec:EFFKONDO}), 
it is seen that, for $1 < \epsilon \le 2$, 
an infinitesimal bosonic coupling $g$ suppresses the Kondo effect.
In other words, the Kondo fixed point is unstable,
and 
this is illustrated in Fig.\,\ref{fig:RG-illustration}(b). 
Our earlier analysis at $\epsilon=2$ is consistent with this general result.
The finding of NRG calculations 
for the BFKM with Ising anisotropy is consistent with our conclusion \cite{Glossop_prb07}.

\sect{Fixed point annihilation and stability of Kondo destruction QCP}
From the above, we can see the following evolution of the fixed point structure. As $\epsilon$ goes from $<1$ to $>1$, 
the critical and Kondo fixed points,
$C'$ and $K$, annihilate with each other. This annihilation happens exactly at $\epsilon=1$,
as illustrated in Fig.\,\ref{fig:RG-illustration}.

This connection with the fixed point annihilation picture leads to another key conclusion of our work. As noted earlier,
the Kondo destruction QCP
 corresponds to the critical fixed point $C'$ at $\epsilon=1^{-}$. 
 At this value of $\epsilon$, the critical fixed point is merging towards the Kondo fixed point $K$. Given that it is zero at the Kondo fixed point, the zero temperature residual entropy likewise vanishes at the Kondo destruction
 QCP. This shows the thermodynamic stability of the QCP.

\sect{Discussion and summary}
Several remarks are in order. First, our work demonstrates the importance of the longitudinal coupling between
the local moment spin and the bosonic bath in its slowly fluctuating regime, $1<\epsilon \le 2$.  
Our analytical results on the SU(2)-symmetric and Ising-anisotropic
Bose-Fermi Kondo model in this 
range of $\epsilon$ are supported by 
the numerical (NRG) results
on the Ising-anisotropic
BFKM \cite{Glossop_prb07}.
This consistency provides
further support for 
the dominant role that the longitudinal fluctuations play 
for this range of $\epsilon$.

Second, Kondo destruction quantum criticality exemplifies correlation-driven localization-delocalization transitions \cite{Paschen-Si_2020,hu_quantum_2024}.
As such, our work contributes to the emerging notion that strange metallicity in diverse correlated systems derives
 from a proximity to this type of transition. By extension, our findings elucidate 
 strongly correlated gapless fermion systems in general.
 
To summarize, we have addressed the nature of Kondo destruction quantum criticality, which has been advanced to describe heavy fermion strange metals. We have linked the Kondo destruction quantum critical point to the general phenomenon of fixed point annihilation; in turn, this uncovers the absence of residual entropy and, thus, the thermodynamic stability of the underlying quantum critical fluid. 
Our work is important to the overall understanding of strange metallicity 
in correlated systems, especially the emerging notion that proximity to 
an electron localization-delocalization transition serves as the underlying mechanism.

We thank R. Bulla, K. Ingersent, M. T. Glossop, S. Paschen,
and M. Vojta for useful discussions.
This work has been supported in part by
the NSF Grant No.\ DMR-2220603, 
and by the Robert A. Welch Foundation Grant No.\ C-1411 
and 
the Vannevar Bush Faculty Fellowship ONR-VB N00014-23-1-2870. 
The majority of the computational calculations have been performed on the Shared University Grid 
at Rice funded by NSF under Grant No.~EIA-0216467, a partnership between Rice University, 
Sun Microsystems, and Sigma Solutions, Inc., 
the Big-Data Private-Cloud Research Cyberinfrastructure MRI-award funded 
by NSF under Grant No. CNS-1338099, and the Advanced Cyberinfrastructure Coordination Ecosystem: 
Services \& Support (ACCESS) by NSF under Grant No. DMR170109.
One of us (J.D.) has been supported by the NSF of China under grant No.\,12274109.
Q.S. acknowledges the hospitality of the Aspen Center for Physics, which is supported by NSF grant No. PHY-2210452.

\bibliography{bfk}

\begin{thebibliography}{62}
\bibitem [{not({\natexlab{a}})}]{note}%
  \BibitemOpen
  \href@noop {} {\bibfield  {journal} {\bibinfo  {journal} {Note the overall
  pre-factor $e^{g^2\beta^2/16}$ in the partition functions,
  Eqs.~(\ref{Z_su2},\ref{Z_Ising}). Its origin lies in the renormalization of
  $Z_b$ by $g$. Some regularization procedure can be introduced to cancel this
  pre-factor. This, however, does not affect the determination of the spin
  responses.}\ } ({\natexlab{a}})}\BibitemShut {NoStop}%
\end{thebibliography}
\bibliographystyle{apsrev4-2}

\setcounter{secnumdepth}{3}

\onecolumngrid
\newpage
\beginsupplement

\section*{Supplemental Materials}
\tableofcontents

 \section{Analytical study for SU(2) Bose-Kondo model 
 }\label{Sec:A} 
 
 In this section, we present the analytical results for the SU(2) Bose-Kondo model at $\epsilon=2$. Without Kondo coupling, 
 \begin{equation}
 \tilde{H}(\vec{\lambda})=hS^z+g\vec{\lambda}\cdot\vec{S}.
 \end{equation}
 Using the parameterization 
$\vec{\lambda}=(\lambda\sin\varphi\cos\theta,
\lambda\sin\varphi\sin\theta,\lambda\cos\varphi)$,
we can specify the eigenvalues of $\tilde{H}(\vec{\lambda})$
as $\pm
g \tilde{\lambda}/2$, with
$\tilde{\lambda}=\lambda\sqrt{(1-y)^2+4y\cos^2\frac{\varphi}{2}}$
and $y=\frac{h}{g\lambda}$. Inserting these into
Eq.\,(\ref{Z_gaussian_averaging}) in the main text leads to 
the local partition function,
$Z_{loc}= Z_b^{-1}Z$:
\begin{eqnarray}
Z_{loc}= \left (2\cosh\frac{h\beta}{2}+
\frac{g^2\beta}{2}\frac{\sinh\frac{h\beta}{2}}{h}
\right )e^{\frac{g^2\beta^2}{16}} \, .
\label{Z_su2}
\end{eqnarray} 
Then the local spin susceptibility is
\begin{eqnarray}
\chi_{loc}=\frac{1}{\beta}\frac{\partial^2 \ln Z_{loc}}{\partial
h^2}|_{h=0}
=\frac{\beta}{12}\frac{3+{g^2\beta^2}/{8}}{1+{g^2\beta^2}/{8}} 
~\,\stackrel{\beta \rightarrow \infty}{\longrightarrow}\,
~\frac{\beta}{12}
\end{eqnarray}
In the low-temperature limit, it has a Curie form,
$\chi_{loc} (T \rightarrow 0) = {\beta}/{12}$,
with a reduced Curie constant
($1/12$ instead of the free-spin value, $1/4$).

The dynamical local spin-spin correlation function,
$\chi_{loc}(\tau) \equiv \langle S^z(\tau)S^z(0)\rangle_{loc}$, 
is obtained exactly from a Gaussian averaging: $\chi_{loc}(\tau) =\left ( \pi^{-3/2} / Z_{loc} \right ) \int d\vec{\lambda}e^{-\vec{\lambda}^2} A(\lambda)$, where $ A(\lambda) =Tr e^{-\beta \tilde{H}[\vec{\lambda}]}S^z(\tau)S^z(0)$.
The trace amounts to  $\sum_{n,m}e^{-E_n\beta}e^{(E_n-E_m)\tau}|\langle n|S^z|m\rangle|^2$, where $n$ and $m$ run over all the eigenstates of $\tilde{H}[\vec{\lambda}]$:
$|+\rangle=
(\cos\frac{\varphi}{2}e^{-i\frac{\theta}{2}},
\sin\frac{\varphi}{2}e^{i\frac{\theta}{2}})$, $|-\rangle=
(-\sin\frac{\varphi}{2}e^{-i\frac{\theta}{2}},
\cos\frac{\varphi}{2}e^{i\frac{\theta}{2}})$.
It follows that
$A(\lambda)
=\frac{1}{2}[\cos^2\varphi\cosh\frac{g\lambda\beta}{2}+
\sin^2\varphi \cosh\frac{g\lambda(\beta-2\tau)}{2}]$
and, in turn,
\begin{eqnarray}
\chi_{loc}(\tau)
=\frac{1}{12}[1+
\frac{2+\frac{g^2(\beta-2\tau)^2}{4}}{1+\frac{g^2\beta^2}{8}}
e^{-\frac{g^2}{4}\tau(\beta-\tau)} ]
~\mathop{\longrightarrow}_{\beta \rightarrow \infty}^{\tau \rightarrow
\beta/2} ~\frac{1}{12} .
\end{eqnarray}
As mentioned in the main text, in the asymptotic low-temperature and long-time limit
($\tau\rightarrow \beta/2$, $\beta \rightarrow \infty$),
it also has a Curie form.

For the
Ising case,
{\it i.e.},
$g_1=g_2=0, g_3=g$, 
the  
partition function is now
\begin{eqnarray}
Z_{loc}=2\cosh\frac{h\beta}{2}e^{\frac{g^2\beta^2}{16}} \, .
\label{Z_Ising}
\end{eqnarray}
The corresponding results for the static and dynamical local spin susceptibilities
are given in the main text.

 \section{Continuous-time quantum Monte Carlo study of SU(2) Bose-Kondo model
 }\label{Sec:B}
The Bose-Kondo model can also be studied numerically using a continuous-time quantum Monte Carlo (CT-QMC) method~\cite{cai2019bose,otsuki2013spin}, for the 
entire range $0<\epsilon<2$.
In CT-QMC, we adopt a different form of $\chi_{0}^{-1}(\tau)$ that is regularized at $\tau = 0$ and $\tau = \beta$ to facilitate numerical computation,
\begin{equation}
\chi_{0}^{-1}(\tau)
=\left[
\frac{\pi/\beta}{\sin(\pi\tau/\beta)} (1+e^{-\beta}-e^{-\tau}-e^{-(\beta-\tau)})
\right]^{2-\epsilon}.
\end{equation}
Here, $ \lim_{\tau \rightarrow 0} \chi_{0}^{-1}(\tau) = \lim_{\tau \rightarrow \beta } \chi_{0}^{-1}(\tau) =1 $. It has the same asymptotic $1/\tau^{2-\epsilon}$ behavior and when $\epsilon=2$, it is identical to the form used in the above exact solution.

 \section{Analytical study of Bose-Fermi Kondo model 
 at $\epsilon = 2$
 }\label{Sec:C}
In this section, we present the details of an exact solution to the BFKM at $\epsilon=2$. 
We treat the Kondo coupling by considering the fermionic Kondo part placed at its Toulouse point \cite{Hewson}. Moreover,
the coupling of the spin to the bosonic bath will be taken to be Ising.

When the Kondo couplings take the Toulouse values, the model
Eq.~(\ref{Z_gaussian_averaging})
can be mapped~\cite{Hewson} to the following non-interacting spinless resonant level model (RLM):
\begin{eqnarray}
H_T(\lambda)=\sum_{\vec{k}} \left [ 
\varepsilon_{\vec{k}}c^{\dagger}_{\vec{k}}
c_{\vec{k}}+V(c^{\dagger}_{\vec{k}}d+h.c.) \right ]
+\varepsilon_d
(d^{\dagger}d-\frac{1}{2}).
\end{eqnarray}
Here, the impurity spin is represented by $S^z\rightarrow d^{\dagger}d-1/2$, $S^+ \rightarrow d^{\dagger}$, and $S^- \rightarrow d$, with $d$ describing a spinless fermion.
The hybridization $V$ is proportional to $J_{K\perp}$, while
$\varepsilon_d=g\lambda+h$.  After integrating out the
fermionic bath, we obtain $ Z_{T}=\pi^{-1/2}\int d\lambda
e^{-\lambda^2}\mathrm{Tr} e^{-\beta H_T(\lambda)} =Z_cZ_{loc}$.
Here, $Z_{c}$ is the partition function of the electron bath,
and $Z_{loc}=\pi^{-1/2}\int d\lambda e^{-\lambda^2}Z_{loc}[\lambda]$
with
\begin{eqnarray}
Z_{loc}[\lambda]=&& 2 
\exp\{-\beta\int_{-D}^{D}
\frac{d\omega}{\pi}n(\omega)
\arctan\frac{\Gamma}{\omega-\epsilon_d}\} \cosh(\frac{\beta\epsilon_d}{2})  ,
\label{Z-RLM}
\end{eqnarray}
where $n(\omega)=[1+\exp(\beta \omega)]^{-1}$ is the Fermi
function, $\Gamma=\pi\rho_0 V^2$ the bare resonance width,
and $\rho_0$ the conduction-electron density of states
at the Fermi energy. Without loss of generality, we assume a flat
band for the electrons, taking the usual limit of a large
bandwidth $D$ ($\gg \Gamma$).

Eq.~(\ref{Z-RLM}) has been derived with care, 
containing not only
the usual phase shift contribution \cite{Hewson},
but also a residual
atomic term. Indeed,  
it  
recovers the results
for both the Ising bosonic Kondo model ($\Gamma=0$)
and the conventional RLM ($g=0$).
When both couplings $\Gamma$ and $g$ are non-zero, the Gaussian
averaging over $\lambda$ complicates the problem.
We focus on
the zero-field static local susceptibility.
It is 
straightforward to show that
\begin{eqnarray}
\chi_{loc}=\frac{\int_{-\infty}^{\infty}d\lambda
e^{-\lambda^2}\chi_{loc}[\lambda] Z_{loc}[\lambda]}
 {\int_{-\infty}^{\infty}d\lambda e^{-\lambda^2}Z_{loc}[\lambda]}|_{h=0},
 \label{chi0}
\end{eqnarray}
where $\chi_{loc}[\lambda]=\{\frac{\partial}{\partial
h}M_{loc}[\lambda]+\beta M_{loc}^2[\lambda]\}$ and
$M_{loc}[\lambda]=\frac{1}{\beta}\frac{\partial \ln
Z_{loc}[\lambda]}{\partial h}$. At low temperatures, 
we use the asymptotic approximation $n(\omega)\approx
\Theta(-\omega)$ (Sommerfeld expansion, valid up to
corrections of order of $T^2/\Gamma$) and integrate over
$\omega$, obtaining 
\begin{equation}
    \chi_{loc}[\lambda]=\frac{1}{\pi\Gamma}\frac{\Gamma^2}
{\Gamma^2+g^2\lambda^2}+\frac{\beta}{\pi^2}\arctan^2\frac{g\lambda}
{\Gamma}
\end{equation}
and 
\begin{equation}
    Z_{loc}[\lambda]=\exp\{\frac{\beta
g\lambda}{\pi}\arctan\frac{g\lambda}{\Gamma}
-\frac{\beta\Gamma}{2\pi}\ln[1+(\frac{g\lambda}{\Gamma})^2]\} Z',
\end{equation}
where $Z'$ is independent of $\lambda$ in the limit
$D/\Gamma\rightarrow\infty$. 

In the weak boson-spin coupling regime, $g/\Gamma \ll 1$, 
one can distinguish two regimes according to the cutoff. 
Within the cutoff $|\lambda| \ll \Lambda$, (referred to as region I),
the dummy variable $\frac{g\lambda}{\Gamma}$ is always small such that expansions like $\arctan\frac{g\lambda}{\Gamma}\approx
\frac{g\lambda}{\Gamma}$ are justified. 
Outside the cut-off ($|\lambda| \gg \Lambda$, region II), $\frac{g\lambda}{\Gamma}$ is large
enough and one has $\arctan\frac{g\lambda}{\Gamma}\approx
\mathrm{sgn} (\lambda)\frac{\pi}{2}$.
$\chi_{loc}[\lambda]$ then becomes $\frac{1}{\pi\Gamma}+\frac{\beta}{\pi^2} \left (\frac{g\lambda}{\Gamma} \right )^2 $ for $|\lambda| \ll \Lambda$, and $\frac{\beta}{4}$ for $|\lambda| \gg \Lambda$. $Z_{loc}[\lambda]$ contains a dimensionless combination,
$\alpha=\frac{\beta g^2}{2\pi\Gamma}$. From this, 
a characteristic temperature scale, $T^*=\frac{
g^2}{2\pi\Gamma}$ (corresponding to $\alpha=1$),
naturally emerges, separating two limiting temperature regimes with distinct asymptotic behavior of $\chi_{loc}$.
At low temperatures, {\it i.e.}, $T\ll T^*$ ( or $\alpha\gg 1$), 
region-II contribution dominates
and $\chi_{loc}\approx\frac{\beta}{4}$. 
In the higher temperature regime,
$T^* \ll T \ll \Gamma$ (or $\alpha\rightarrow 0$), 
the region-I contribution dominates,
yielding  
$\chi_{loc}\approx\frac{1}{\pi\Gamma}$. A
smooth but broad crossover occurs 
around $T \sim T^*$.

\section{Suppression of the Kondo effect by singular bosonic fluctuations}\label{Sec:EFFKONDO}

In this section, we start from the SU(2) Bose--Fermi Anderson model and perform a Firsov--Lang transformation followed by a Schrieffer--Wolff transformation to obtain an effective Kondo model \cite{Pixley_2011}. We then show that for the case of a singular bosonic bath, $1 < \epsilon \le 2$, the Kondo effect
is suppressed by an arbitrarily small bosonic coupling $g$.

The SU(2) Bose-Fermi Anderson model is written as:
\begin{align}
    H = H_{F}+H_{B}+\sum_{\sigma}\epsilon_{d}d^{\dagger}_{\sigma}d_{\sigma}+\sum_{k,\sigma}(V_{k}d^{\dagger}_{\sigma}c_{k\sigma}+h.c) +Un_{d\uparrow}n_{d\downarrow}+g\sum_{a=1}^{3}S^{a}\Phi^{a}
\end{align}
where  $H_F=\sum_{k,\sigma}\varepsilon_{k}c^{\dagger}_ {k\sigma}
c_{k\sigma}$ and 
$H_B=\sum_{p,a}w_{p}\Phi^{\dagger}_{pa}\Phi_{pa}$
respectively describes a conduction-electron band 
and a bosonic bath, with $\varepsilon_k$ and $w_{p}$ denoting 
the corresponding dispersions. $\sigma=\uparrow,\downarrow$ marks the spin quantum number,
and $a=1,2,3$ the components of the vector boson field. $\Phi^a \equiv 
\sum_{p}(\Phi_{pa}+\Phi^{\dagger}_{-pa})$ is the real bosonic field that 
couples with the spin of the localized $d$-electrons: $S^{a}=\frac{1}{2}d^{\dagger}_{\sigma}\tau^{a}_{\sigma\sigma'}d_{\sigma'}$, where $\tau^{a}$ is the Pauli matrix.

\subsection*{1. Firsov--Lang (FL) transformation}

To remove the $S^z\Phi^z$ term we choose
\begin{equation}
S_{\rm FL} = g S^z \sum_q \frac{1}{\omega_q}\big(\Phi_q^{z\dagger} - \Phi_{q}^{z}\big),
\end{equation}
and define $\tilde H = e^{S_{\rm FL}} H e^{-S_{\rm FL}}$.  
The localized-electron operators are dressed as
\begin{equation}
\tilde d_\sigma = e^{S_{\rm FL}} d_\sigma e^{-S_{\rm FL}}
= d_\sigma\,\exp\!\left\{\frac{\sigma g}{2}\sum_q\frac{1}{\omega_q}(\Phi_q^{z\dagger}-\Phi_{q}^{z})\right\},
\qquad \sigma=\pm1,
\label{eq:S2}
\end{equation}
and the z-component boson operator is shifted as:
\begin{align}
    e^{S_{FL}}\Phi_{q}^{z}e^{-S_{FL}}=\Phi_{q}^{z}-gS^{z}/\omega_{q},
\end{align}
so that the transformed Hamiltonian reads
\begin{align}
\tilde H &= H_{F} + H_{B} + \tilde H_{\rm loc} + \tilde H_{\rm hyb} + H_{\perp}, \\
H_{\rm B} &= \sum_q \omega_q \Phi_q^\dagger\Phi_q, \\
\tilde H_{\rm loc} &= \tilde\epsilon_d (n_{d\uparrow}+n_{d\downarrow}) + \tilde U\, n_{d\uparrow}n_{d\downarrow}, \\
\tilde H_{\rm hyb} &= \sum_{k,\sigma}\!\left(V_k c_{k\sigma}^\dagger \tilde d_\sigma + \text{h.c.}\right),
\end{align}
with renormalized parameters
\begin{equation}
\tilde U = U + \tfrac{1}{2} g^2\sum_q \frac{1}{\omega_q},\qquad
\tilde\epsilon_d = -\frac{\tilde U}{2}.
\label{eq:S4}
\end{equation}
Here $H_\perp$ denotes the residual transverse boson couplings
$g(S^x\Phi^x+S^y\Phi^y)$: $H_{\perp}=g/\sqrt{2}(\tilde{S}^{+}\Phi^{-}+\tilde{S}^{-}\Phi^{+})$, where $\Phi^{\pm}=\frac{1}{\sqrt{2}}(\Phi^{x}\pm i\Phi^{y})$, and $\widetilde{S}^{+}=d^{\dagger}_{\uparrow}d_{\downarrow}\exp\left\{g\sum_{q}\frac{1}{\omega_{q}}(\Phi_q^{z\dagger}-\Phi_{q}^{z})\right\},\widetilde{S}^{-}=d^{\dagger}_{\downarrow}d_{\uparrow}\exp\left\{-g\sum_{q}\frac{1}{\omega_{q}}(\Phi_q^{z\dagger}-\Phi_{q}^{z})\right\}$ are dressed spin-flip operators.

\subsection*{2. Schrieffer--Wolff (SW) transformation}
 
For the regime of sufficiently large $U$, we perform a SW transformation with generator
\begin{equation}
S_{\rm SW} = \sum_{k,\sigma} V_k
\left(\frac{1-n_{d,-\sigma}}{\tilde\epsilon_d-\varepsilon_k}
+\frac{n_{d,\sigma}}{\tilde\epsilon_d+\tilde U-\varepsilon_k}\right)
\big(\tilde d_\sigma^\dagger c_{k\sigma}-c_{k\sigma}^\dagger \tilde d_\sigma\big).
\end{equation}
Projecting out empty and doubly occupied local-electron states yields the effective Kondo Hamiltonian
\begin{align}
H_{\rm eff} &= \sum_{k,\sigma}\varepsilon_k c_{k\sigma}^\dagger c_{k\sigma}
+ \sum_q \omega_q \Phi_q^\dagger\Phi_q
+ \sum_{k,k',\sigma}\Big(\tfrac{1}{2}\widetilde W_{kk'}+\tfrac{1}{4}\widetilde J_{kk'}\Big)c_{k\sigma}^\dagger c_{k'\sigma} \nonumber\\
&\quad - \sum_{k,k'} \widetilde J_{kk'}\left\{
\tfrac{1}{2}\big(s_{k}^+ \widetilde S^- + s_{k}^- \widetilde S^+\big) + s_{k}^z S^z\right\}+\widetilde{H}_{\perp},
\label{eq:S6}
\end{align}
with conduction-electron spin densities $s^\alpha_k$ and dressed local-electron spin operators
$
\widetilde S^+ = S^+ \exp\!\Big\{ g\sum_q\frac{1}{\omega_q}(\Phi_q^{z\dagger}-\Phi_{q}^{z})\Big\},
\widetilde S^- = S^- \exp\!\Big\{-g\sum_q\frac{1}{\omega_q}(\Phi_q^{z\dagger}-\Phi_{q}^{z})\Big\}.
$ $\widetilde{H}_{\perp}=e^{S_{SW}}H_{\perp}e^{-S_{SW}}=H_{\perp}+\frac{1}{\sqrt{2}}\sum_{k}\widetilde{F}_{k}[d_{\uparrow}^{\dagger}c_{k\downarrow}n_{d\downarrow}-c_{k\uparrow}^{\dagger}d_{\downarrow}(1-n_{d\uparrow})]\Phi^{-}e^{\frac{3}{2}g^2\sum_{q}\frac{1}{\omega_{q}}(\Phi^{z\dagger}_{q}-\Phi_{q}^{z})}+h.c+O(V^2/\widetilde{U}^2)$.
The $\widetilde{J}_{kk'},\widetilde{W}_{kk'},\widetilde{F}_{k}$ couplings are defined as
\begin{align}
\widetilde J_{kk'} &= V_k V_{k'}\left(
\frac{1}{\varepsilon_k-\tilde\epsilon_d}+\frac{1}{\varepsilon_{k'}-\tilde\epsilon_d}
-\frac{1}{\varepsilon_k-\tilde\epsilon_d-\tilde U}-\frac{1}{\varepsilon_{k'}-\tilde\epsilon_d-\tilde U}
\right), \label{W}\\
\widetilde W_{kk'} &= V_k V_{k'}\left(\frac{1}{\varepsilon_k-\tilde\epsilon_d}+\frac{1}{\varepsilon_{k'}-\tilde\epsilon_d}\right),\label{J}\\
\widetilde{F}_{k}&=\frac{gV_{k}}{\tilde{\epsilon}_{d}-\epsilon_{k}}.\label{F}
\end{align}

\subsection*{3. The case of singular bosonic bath}
The FL transformation renormalizes $\tilde U$ and $\tilde\epsilon_d$ and dresses the local-electron operators with bosonic displacement factors.  
The subsequent SW projection leads to a Bose--Fermi Kondo model in which the spin-flip operators $\widetilde S^\pm$ acquire exponentials of bosonic displacements.  
The interaction shift is
\begin{align}
    \widetilde U - U
    = \tfrac{1}{2} g^{2}\sum_{q}\frac{1}{\omega_{q}}
    = \tfrac{g^2}{2}\int_{0}^{\infty}\! d\omega\,\frac{J(\omega)}{\omega}
    \;\propto\; \tfrac{g^2}{2}\int_{0}^{\Lambda}\! d\omega\,\frac{1}{\omega^{\epsilon}} ,
\end{align}
which diverges for $1 < \epsilon \le 2$ and so does for $\tilde{\epsilon}_{d}$.  
As a result, the effective couplings, $\widetilde J$,  $\widetilde W$ and $\widetilde{F}$, vanish.  
Hence the system flows away from the Kondo fixed point 
for nonzero but arbitrarily small $g$ and 
towards the strong-coupling local moment fixed point.
This is in accordance with Fig.~\ref{fig:RG-illustration}(b).

\end{document}